# Range Improvement in Single-Beam Phased Array Radars by Amplifier Impedance Tuning

Pedro Rodriguez-Garcia, Jack Sifri, Caleb Calabrese, Charles Baylis, and Robert J. Marks II

*Abstract*—**Significant radar range degradation can be incurred due to variations in antenna impedance from changing array scan angle. Element-wise reconfigurable impedance tuners can be used to optimally match the power amplifier device; however, the impedance matching can also affect the array pattern. In this paper, the effects of element-wise impedance tuning on the transmitted power and on the array pattern are studied, and an approach is recommended for the element-wise implementation of impedance tuners. Examples of impact on array pattern and transmitted power are given using simulation of a designed switched-stub impedance tuner. As a result of these experiments, guidelines are developed for the creation of real-time circuit optimization techniques in the array elements.**

## I. INTRODUCTION

IN a radar phased array system, mutual coupling between antenna elements can cause significant variations in impedance presented by the transmit antennas to the preceding circuitry when the scan angle varies [1]. This can negatively impact radar range capabilities. Reconfigurable circuitry can be used to provide matching to a power amplifier for improvement of output power, power-added efficiency (PAE), and/or spectral performance [2], and therefore can improve range. We have shown in a simulation example that presenting the power amplifier with an antenna impedance away from the antenna impedance for which an amplifier load matching network was designed could reduce the output power by approximately 3 dB, which translated to a communication range reduction of 29.3%, or a radar range reduction of 15.9% [3]. To avoid such reductions, impedance tuners can be placed in each array element. For radar, real-time impedance tuning has long been impractical due to a lack of high-power tunable circuitry. Semnani, however, demonstrates a 90 W evanescent-mode S-band cavity tuner design that could conceivably be used in radar phased array tuning [4]. The works in [5]-[7] describe several reconfigurable antenna impedance tuning methodologies that adapt for changes in single-element antenna impedances due to environmental changes such as human hand and body effects in wireless communications handsets. An expansion is needed for a multi-antenna phased array system, as well as an analysis in the changing scan angles and transmission antenna gain for an active electronically scanned array (AESA). Several other works have shown that wide-angle array impedance matching can be achieved [8]-[11] using various design techniques inherent to the phased array itself to resolve the array impedance mismatch. Implementing element-wise reconfigurable matching networks, however, is expected to provide more output power capability and greater flexibility in a real-time system than static wide-angle matched designs. Rather than a typical fixed power-amplifier matching network to match the antenna impedance to the optimal power-amplifier load impedance, a tunable matching network can allow the amplifier to be consistently loaded with its optimum termination as the array scan angle is varied, without degrading the array pattern. We address the array impedance variation issue by implementing individual tuning of the driven element impedances in a phased array transmitter according to the Fig. 1 diagram as the scan angle changes. Through a co-simulation setup between Keysight Technologies Advanced Design System (ADS) circuit simulation software and ADS Momentum electromagnetic simulation software, we analyze the effect of element-wise reconfigurable circuitry on the range performance of radar, as well as on the transmitting array pattern. The results of this work are also expected to be applicable to fifth-generation (5G) wireless communication systems utilizing phased arrays.

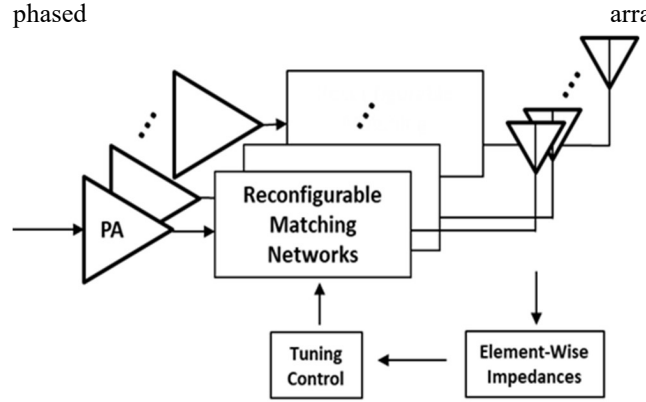

Fig. 1. Block diagram of element-wise array matching circuit configuration

## II. MUTUAL COUPLING IN PHASED ARRAY TRANSMITTERS

Mutual coupling, according to Haupt [12], is the interaction between an antenna and its environment. In a phased array transmitter, an antenna element interacts with the other antenna elements in the array. The radiation from one antenna element causes apparent reflections in the other antenna elements since every element receives some of the radiation from the other elements. Fig. 2 illustrates this mutual coupling effect in a simple two-element array. The second antenna radiates a wave that is received by the first antenna and vice versa. The received waves at each element travel from the antennas back to their sources, resembling reflected waves, causing changes in the driven element impedances presented by the antennas. As the scan angle is varied, the amount of mutual coupling between elements changes, changing the driven element impedances.

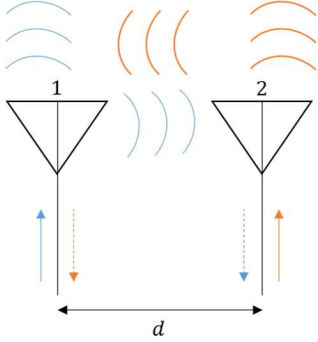

Fig. 2. Radiation representation of mutual coupling between two phased array antenna elements

The self and mutual impedances of each antenna element in a phased array can be represented by the well-known impedance matrix $[Z]$, which relates currents and voltages of the elements:

$$\begin{bmatrix} V_1 \\ \vdots \\ V_N \end{bmatrix} = \begin{bmatrix} Z_{11} & \cdots & Z_{1N} \\ \vdots & \ddots & \vdots \\ Z_{N1} & \cdots & Z_{NN} \end{bmatrix} \begin{bmatrix} I_1 \\ \vdots \\ I_N \end{bmatrix}. \quad (1)$$

For a two-element array, (1) reduces to the following:

$$\begin{aligned} V_1 &= Z_{11}I_1 + Z_{12}I_2 \\ V_2 &= Z_{21}I_1 + Z_{22}I_2. \end{aligned} \quad (2)$$

Assuming that the two-element array is a two-port linear circuit network, the self-impedance of each antenna is the impedance of the isolated antenna in the absence of the other antenna. If the current flowing into element 2 is zero (this can be enforced by terminating element 2 in an open circuit), the ratio $V_1/I_1$ is the $Z_{11}$ term. Similarly, if antenna 1 is terminated in an open circuit, then the impedance presented by antenna 2 to its transmitter circuit is the self-impedance $Z_{22}$:

$$Z_{11} = \frac{V_1}{I_1}\Big|_{I_2=0} \quad Z_{22} = \frac{V_2}{I_2}\Big|_{I_1=0}. \quad (3)$$

The mutual impedances between antenna elements describe the mutual coupling. These terms represent the open-circuit voltage across an antenna input that results from a current input to the other antenna. For the two-antenna system, the mutual impedance terms are defined as follows based on (2):

$$Z_{12} = \frac{V_1}{I_2}\Big|_{I_1=0} \quad Z_{21} = \frac{V_2}{I_1}\Big|_{I_2=0}. \quad (4)$$

When antennas are placed in an array and excited simultaneously, the "driven element impedance" is the ratio of voltage across the antenna port to the current entering the antenna. The driven element impedances are a sum of their self and mutual impedances:

$$Z_{d1} = \frac{V_1}{I_1} = Z_{11} + Z_{12}\frac{I_2}{I_1}$$

$$Z_{d2} = \frac{V_2}{I_2} = Z_{21}\frac{I_1}{I_2} + Z_{22}. \quad (5)$$

The driven element impedance equations in (5) can be expanded to compensate for any amount of array elements as well as for changing scan angle. For a linear two-element array, zero phase can be assumed for the first antenna current, and the second antenna current is assigned the same magnitude, with a phase shift based on the scan angle $\theta_s$:

$$\begin{aligned} I_1 &= |I_1| \\ I_2 &= |I_2|e^{-jk_0 d \sin\theta_s}. \end{aligned}$$

For a linear array configuration, the generalized driven element impedance an $N$-element array with scan angle $\theta_s$ is given as follows [12]:

$$Z_{dn} = \sum_{m=1}^{N} Z_{nm} \frac{|I_m|}{|I_n|} e^{jk_0(n-m)d\sin\theta_s}. \quad (6)$$

From (6), the self-impedance terms $(n = m)$ remain unchanged. Steering the array, however, changes the mutual impedance terms $(n \neq m)$. Driven element impedance mismatches can therefore occur as a result of changing the array scan angle and can be mitigated properly with reconfigurable element-wise impedance tuners. Precautions must be taken, however, when tuning the individual array elements. When individual driven element impedances are altered in equation (6), the current sources are also altered in magnitude and phase. To preserve the integrity of the antenna pattern, the relative phases and amplitudes of the individual elements needed to steer the array to the desired scan angle must be unharmed. Although driven impedances for different array elements may be different, tuning to match each element individually can cause distortion in the beam pattern in such cases. Impedance tuning should therefore be performed to maximize power, while ensuring the actual transmitted beam pattern is close enough to the canonical beam pattern to be considered acceptable.

## III. Phased Array Impedance Tuning

Simulations were conducted using the ADS circuit and Momentum EM simulators. A schematic was first generated with a uniform linear array (ULA) of four $\lambda/2$-spaced microstrip patch antenna elements (Fig. 3). The elements were designed using Rogers RO4003C substrate at the design frequency of 3.55 GHz, a frequency presently allocated for sharing between radar and wireless communications in the United States. Nonlinear models for the MWT-173 Gallium Arsenide (GaAs) metal-semiconductor field-effect transistor (MESFET) biased at $V_{DS} = 4.5$ V and $V_{GS} = $ -1.5 V were connected to the antenna elements.

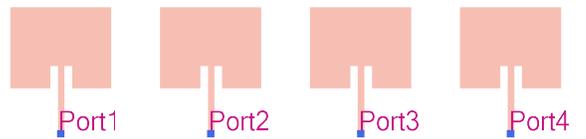

Fig. 3. 4-element $\lambda/2$ uniform linear microstrip array in ADS.

Fig. 4 shows a parameter sweep of the driven element impedances at 3.55 GHz as the array scans from $\theta_s = -60°$ to $+60°$ in the $\phi = 0°$ cut that was conducted to observe the mismatch effects of varying the scan angle. In Fig. 4, the driven end element impedances (elements 1 and 4) behave similarly and the driven inner element impedances (elements 2 and 3) behave similarly because of array symmetry.

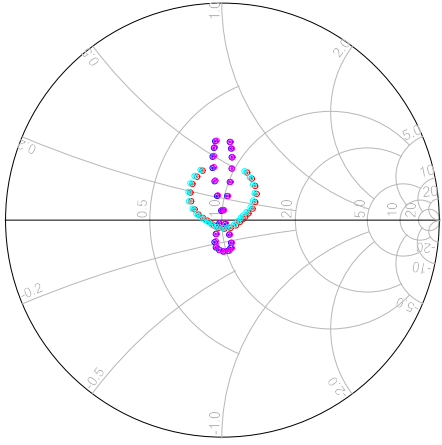

Fig. 4. Parameter swept driven element impedances for scan angles of $\theta_s = -60°$ to $+60°$ at 3.55 GHz for $k_0 = 2\pi/\lambda$, $d = \lambda/2$, end elements (light blue and red), and inner elements (dark blue and purple).

The driven element impedances for this array of four elements follow from equation (6):

$$Z_{d1} = \frac{V_1}{I_1} = Z_{11} + Z_{12}\frac{I_2}{I_1} + Z_{13}\frac{I_3}{I_1} + Z_{14}\frac{I_4}{I_1}$$
$$Z_{d2} = \frac{V_2}{I_2} = Z_{21}\frac{I_1}{I_2} + Z_{22} + Z_{23}\frac{I_3}{I_2} + Z_{24}\frac{I_4}{I_2}$$
$$Z_{d3} = \frac{V_3}{I_3} = Z_{31}\frac{I_1}{I_3} + Z_{32}\frac{I_2}{I_3} + Z_{33} + Z_{34}\frac{I_4}{I_3}$$
$$Z_{d4} = \frac{V_4}{I_4} = Z_{41}\frac{I_1}{I_4} + Z_{42}\frac{I_2}{I_4} + Z_{43}\frac{I_3}{I_4} + Z_{44}. \quad (7)$$

To steer the beam to a direction $\theta_s$, the individual current sources in (7), for a uniformly spaced linear array, follow the linear phase progression

$$\begin{array}{ll} I_1 = |I_1| & I_2 = |I_2|e^{-jk_0 d\sin\theta_s}, \\ I_3 = |I_3|e^{-j2k_0 d\sin\theta_s} & I_4 = |I_4|e^{-j3k_0 d\sin\theta_s}. \end{array} \quad (8)$$

It is imperative that this phase progression remain unharmed when tuning the driven element impedances in (7) to preserve the integrity of the array pattern. From (8), the first element is typically excited with no phase shift. The next consecutive elements all experience varying phase shifts so these relative phase shifts should be maintained to provide an undistorted array pattern and steer the array properly to the desired scan angle.

### A. Tuning at Broadside

Upon observing these effects, the array was then steered to the broadside ($\theta_s = 0°$) scan angle with identical magnitude and phase current source excitations of each array element. A power delivered load-pull was then conducted to observe the optimum terminating load impedance that should presented to the amplifier for maximum power delivery. The driven impedances were placed on the same load-pull plot in Fig. 5 to show the power delivered load impedance level in which the un-tuned driven element impedances were located as well as the array pattern. Due to the equality of all four excitation currents, the initial driven element impedances for the broadside case are represented as a special case of equation (7):

$$Z_{d1} = \frac{V_1}{I_1} = Z_{11} + Z_{12} + Z_{13} + Z_{14}$$
$$Z_{d2} = \frac{V_2}{I_2} = Z_{21} + Z_{22} + Z_{23} + Z_{24}$$
$$Z_{d3} = \frac{V_3}{I_3} = Z_{31} + Z_{32} + Z_{33} + Z_{34}$$
$$Z_{d4} = \frac{V_4}{I_4} = Z_{41} + Z_{42} + Z_{43} + Z_{44}. \quad (9)$$

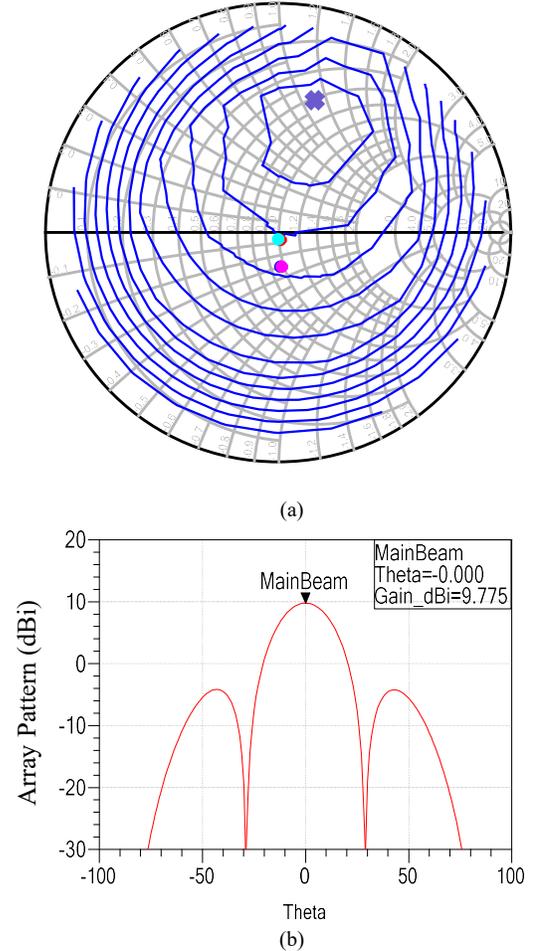

(a)

(b)

Fig. 5. (a) Untuned driven element impedances for $\theta_s = 0°$, (b) untuned array pattern (dBi).

Fig. 5(a) shows the un-tuned driven element impedances with the output power load-pull contours. The un-tuned driven element impedances are not located at the maximum output power load impedance and according to the radar range equation,

$$R_{max} = \left[ \frac{P_t G^2 \lambda^2 \sigma}{(4\pi)^3 L \left( \frac{S}{N} \right)_{min} k_b T_s B} \right]^{\frac{1}{4}}, \qquad (10)$$

this power deficiency can result in a significant reduction of range detection radar capabilities if the driven element impedances are left unchanged without impedance tuning.

To mitigate this reduction of radar range and maximize the transmitter output power, the driven element impedances are tuned with models for the switched-state radial stub reconfigurable impedance tuners presented by Calabrese [13]. A tuner is placed between each antenna element and its associated power amplifier. The layout of the tuner and its 3.55 GHz driven element impedance coverage at broadside for each element is presented in Fig. 6.

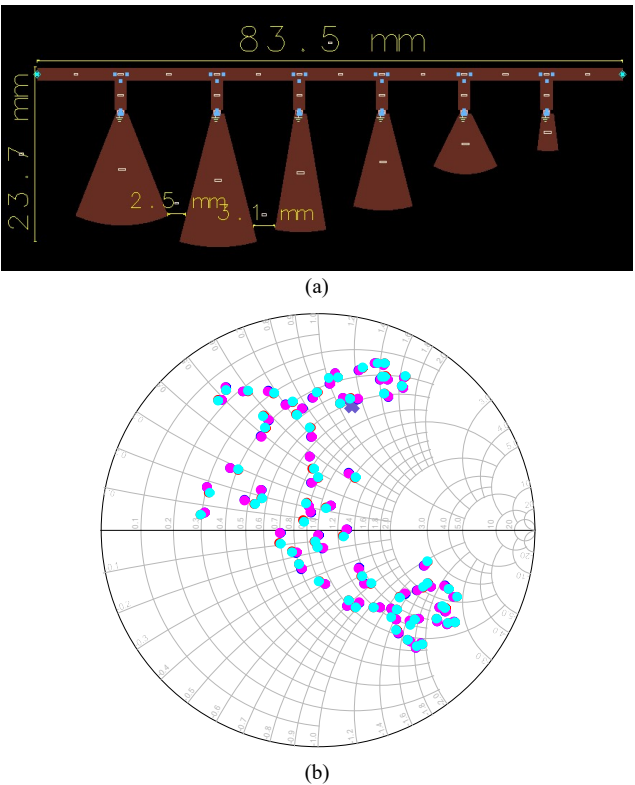

(a)

(b)

Fig. 6. (a) Board layout of the switched-state radial stub impedance tuner, (b) tuner element-wise driven impedance coverage for $\theta_s = 0°$ showing closest points to maximum PA power delivered load impedance (purple x-mark) at switch state 111110 for all tuners.

The tuner switch state combination can be represented with a binary sequence. A bit of "1" indicates a radial stub has been activated (presented to the series line by closing the switch), and a bit of "0" indicates a radial stub has not been activated (Fig. 6(a)). Since there are a total of six radial stubs in the design, this switched-stub configuration provides a total combination of $2^6$ or 64 unique tuning states. The 3.55 GHz impedance coverage (Fig. 6(b)) shows that it is possible to tune the driven impedances from Fig. 5(a) for the $\theta_s = 0°$ to the optimum power delivered load impedance to be presented to each power amplifier using this reconfigurable impedance tuner design. Tuner losses at the varying switch states, however, may differ significantly since different stubs are "activated" and "deactivated" according to the switch state. This means that not only must the impedance tuning mechanism consider the maximum power delivered load impedance, but also the tuner loss at the different switch states. The tuner design is discussed in greater detail in [13].

From Fig. 6(b), there are at least two tuner switch states that could match the driven element impedances to the optimum PA load impedance, however, the losses may be different for the two states, so it is desired to find the state that provides the optimum tuner output power considering state-wise tuner losses. For this reason, the fast tuning algorithm presented by Calabrese [13] is used to determine the switch state needed to tune all driven element impedances to the maximum output power and account for tuner losses. The algorithm begins with the switch state of 000000 (all switches off). The power is subsequently measured at the output of each tuner which is fed to each element in the array. The first switch is toggled to the activated state and the output power is measured for improvement. If output power is improved by activation of the first switch, that switch remains activated and the process iterates through the rest of the switches. The algorithm is then re-iterated until no further improvement in output power is accomplished and the switch state that gives the maximum output power is selected [13]. In performing the search algorithm for driven impedances resulting from $\theta_s = 0°$, the maximum output power is accomplished using a switch state sequence of 111110. This tuning state is applied to the tuners in all four elements to preserve the linear phase progression of the current sources; maintaining the scan angle and the array pattern shape, and tuning the driven element impedances to the optimum power delivered PA load impedance. The resulting tuned driven element impedances and the resulting array pattern for $\theta_s = 0°$ are shown in Fig. 7. Fig. 7(a) shows that all tuned impedances presented to the power amplifiers are near the target load impedance on the Smith Chart.

As a result of the impedance matching, the achievable resulting transmit power increases from 19.2 dBm to 22.1 dBm, providing 18.2% radar range improvement. Fig. 7(b) shows that the array scan angle and overall beam shape remain unharmed. Since each element uses the same tuner switch state of 111110, the adjustments to the magnitude and phase of the transmitted waveform in all elements are expected to be identical, preserving the beam pattern.

Fig. 8 shows that tuning inner and outer elements differently can result in undesirable effects on the array pattern. The inner elements were tuned using the switch state sequence of 111110 whereas the outer elements were tuned with the switch state sequence of 000010, which is the state providing the next closest $\Gamma_L$ to the maximum PA power delivered $\Gamma_L$ considering tuner losses. Because similar values of $\Gamma_L$ are presented in all elements, similar impedance matching is achieved. However, the number of exposed stubs is very different between the two states, causing significant magnitude and phase adjustments to the voltage at the output of the tuner. Removing magnitude and/or phase equality between the circuitry in the different transmitter elements causes the transmitted pattern to be altered, as shown in Fig. 8(b). The antenna gain at the intended scan angle of $\theta_s = 0°$ is reduced from 9.767 dBi to -0.801 dBi (Fig. 9(b)), and the two sidelobes possess higher gain than the desired scan angle. To steer the array to $\theta_s = 0°$, the phases of all current excitations for all elements must be identical. Tuning the outer two elements with a different switch state alters their currents to be out of phase with the inner elements. Fig. 9(a) shows that all four elements have the same transmission phase when the matching networks are tuned to the same switch setting, whereas Fig. 9(b) shows that two different transmission phases exist for the scenario where the outer and inner elements have different matching network switch settings. The magnitude and phase of $S_{21}$ are shown in Tables I and II, respectively, for each of the four elements. The difference in $S_{21}$ magnitudes in Fig. 9(a) is small at 3.55 GHz, but the difference in phases between the inner and outer elements is approximately 143°. As such, the phase differences between inner and outer element transmissions seems to be the main cause of the beam distortion visible in Fig. 8(b), when compared to the case where all elements are tuned identically (Fig. 7(b)).

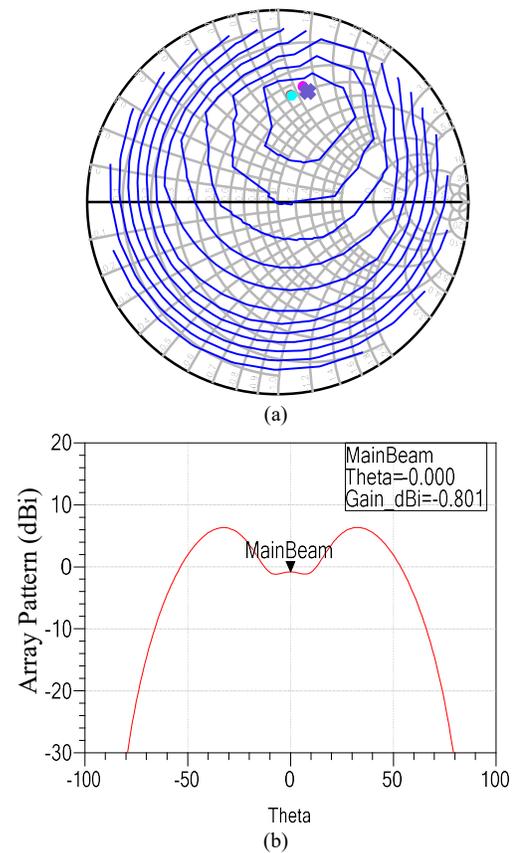

(a)

(b)

Fig. 8. (a) Non-identically tuned driven element impedances for $\theta_s = 0°$, (b) Non-identically tuned array gain pattern (dBi).

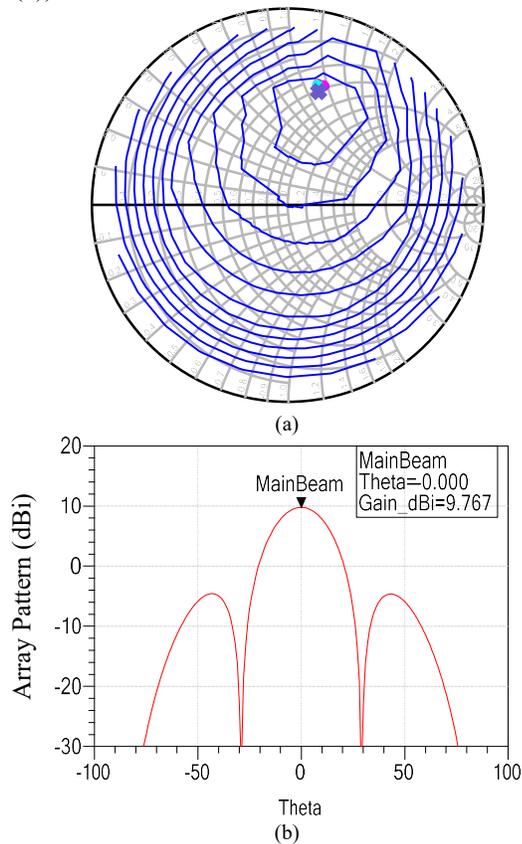

(a)

(b)

Fig. 7. (a) Tuned driven element impedances for $\theta_s = 0°$, (b) identically tuned array gain pattern (dBi).

TABLE I
TUNER TRANSMISSION COEFFICIENTS IDENTICALLY TUNED

| Element | $|S_{21}|$ | $\angle S_{21}$ |
|---|---|---|
| 1 | 0.54 | 38.78° |
| 2 | 0.54 | 38.78° |
| 3 | 0.54 | 38.78° |
| 4 | 0.54 | 38.78° |

TABLE II
TUNER TRANSMISSION COEFFICIENTS NON-IDENTICALLY TUNED

| Element | $|S_{21}|$ | $\angle S_{21}$ |
|---|---|---|
| 1 | 0.49 | -105.08° |
| 2 | 0.54 | 38.78° |
| 3 | 0.54 | 38.78° |
| 4 | 0.49 | -105.08° |

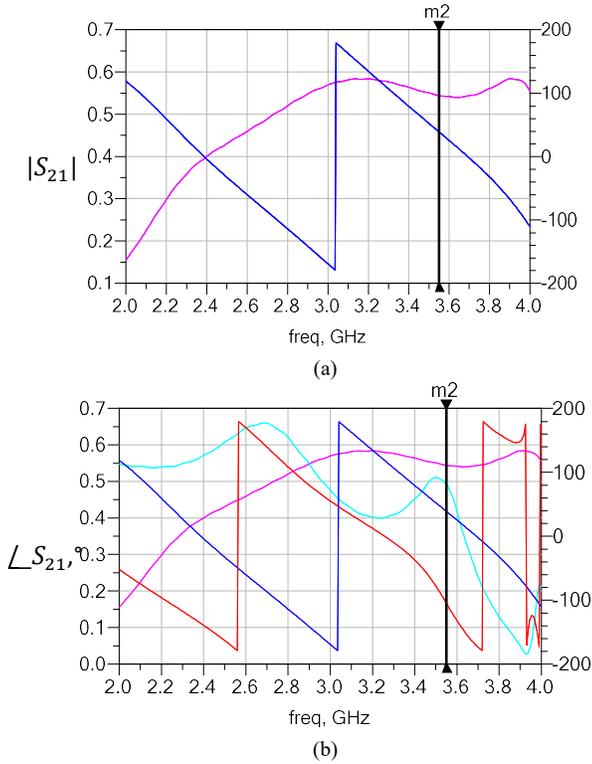

(a)

(b)

Fig. 9. (a) Magnitudes (pink, y-axis) and phases (dark blue, right y-axis) of $S_{21}$ of each tuner when tuned with all same switch states, (b) Magnitudes (purple, y-axis) of inner elements, magnitudes (light blue, y-axis) of end elements, phases (dark blue, right y-axis) of inner elements, and phases (red, right y-axis) of $S_{21}$ of each tuner when only inner and outer elements tuned with the same switch states.

When all elements are not identically tuned, the phase progression can be disrupted, and the array pattern shape will be distorted. The same principle applies across all scan angles.

### B. Tuning at Varying Scan Angles

As the array scan angle $\theta_s$ deviates from broadside, the mutual coupling effects change, and the initial driven element impedances will also change, as calculated by equation (6). Figs. 10-13 show the initial un-tuned driven element impedances and array patterns of the four elements when the array is steered to the scan angles of $\theta_s = -60°$, $-25°$, $+10°$, and $+30°$, respectively. The differences between driven impedances of different elements are greater in these scenarios.

To increase radar detection range capabilities for $\theta_s = +60°$, the driven element impedances from Fig. 10 must once again be tuned using the algorithm in [13] to provide the maximum output power considering tuner losses. The switched-state stub tuner's load impedance coverage for each of the four scan angles (resulting in different driven impedances) at 3.55 GHz is shown in Fig. 14 with the PA maximum-power load impedance. The tuner setting to match the initial driven impedance to the desired PA load impedance while minimizing tuner loss is desired.

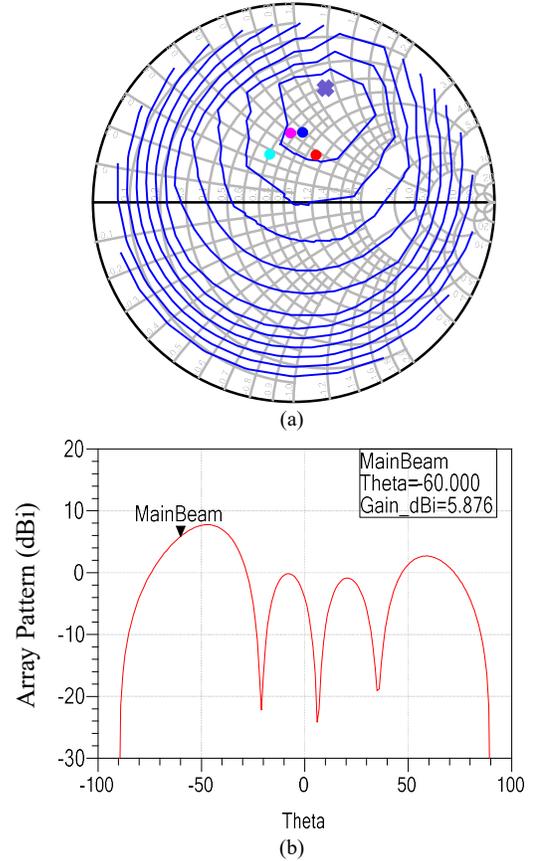

(a)

(b)

Fig. 10. (a) Untuned driven element impedances for $\theta_s = -60°$, (b) untuned array pattern (dBi).

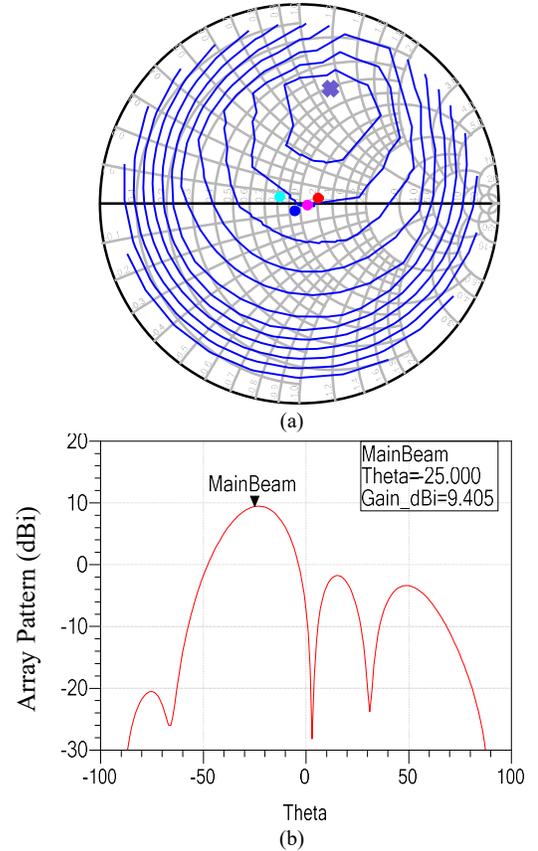

(a)

(b)

Fig. 11. (a) Untuned driven element impedances for $\theta_s = -25°$, (b) untuned array pattern (dBi).

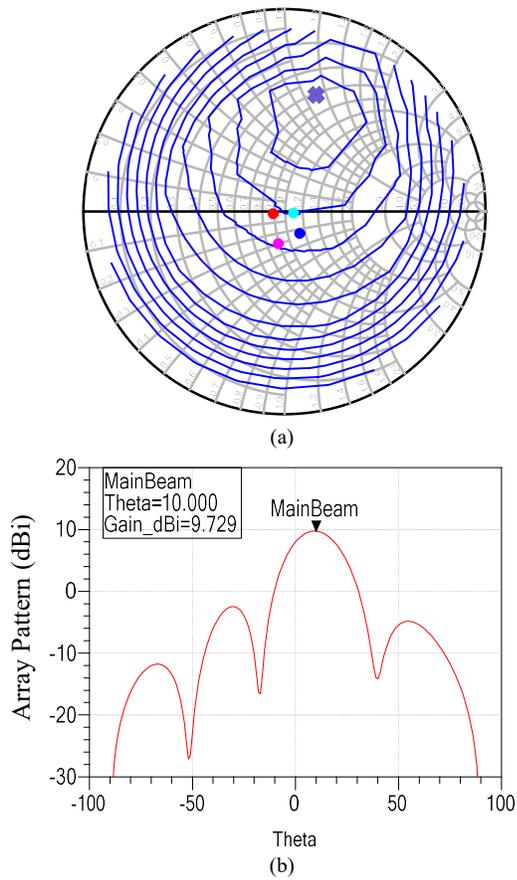

Fig. 12. (a) Untuned driven element impedances for $\theta_s = +10°$, (b) untuned array pattern (dBi).

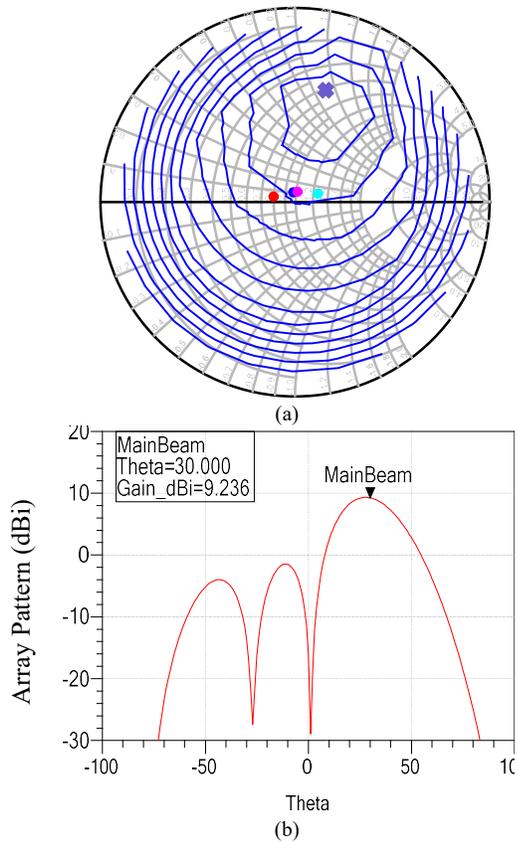

Fig. 13. (a) Untuned driven element impedances for $\theta_s = +30°$, (b) untuned array pattern (dBi).

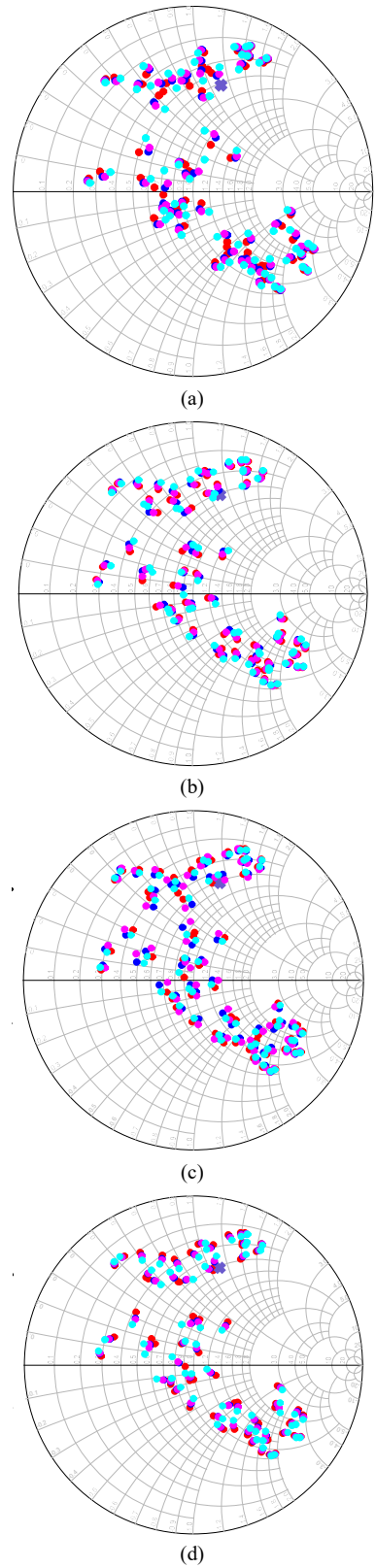

Fig. 14. (a) Tuner element-wise driven impedance coverages (end elements = light blue and red, inner elements = dark blue and purple) for $\theta_s = -60°$, (b) $\theta_s = -25°$, (c) $\theta_s = +10°$, and (d) $\theta_s = +30°$. The amplifier maximum-power load impedance is labeled with an 'X'.

Implementing the iterative algorithm in [13], the switch state 111110 was once again determined to be the optimal switch state for all scan angles. Using this tuner configuration for all scan angles, all driven element impedances were tuned with this switch state on each element to preserve the scan angle and array pattern shape. Calculated range improvements are shown in Table III for the different scan angles; the improvements range from 18% to nearly 27%. Figs. 15-18 show how well the switched-state tuners are able to tune all driven element impedances to the desired PA load impedance at each varying scan angle. The results show the tuners are robust enough to overcome the mutual couplings of the elements at the varying scan angles, regardless of the initial asymmetry of the driven impedances on the Smith Chart. Additionally, the array pattern integrity is preserved, as seen by comparing the tuned patterns (Figs. 15(b)-18(b)) with the untuned patterns (Figs. 10(b)-13(b)), since all elements at each scan angle were tuned using the same switch state. The relative phases of the antenna input signals, therefore, remain the same.

TABLE III
ACHIEVABLE TUNED RADAR RANGE IMPROVEMENT

| SCAN ANGLE | AVG. INITIAL POWER | AVG. TUNED POWER | RADAR RANGE IMPROVEMENT |
|---|---|---|---|
| -60° | 17.2 dBm | 21.9 dBm | +26.5% |
| -25° | 18.6 dBm | 21.9 dBm | +21.5% |
| +10° | 19.1 dBm | 22.1 dBm | +18.9% |
| +30° | 18.3 dBm | 22.0 dBm | +23.7% |

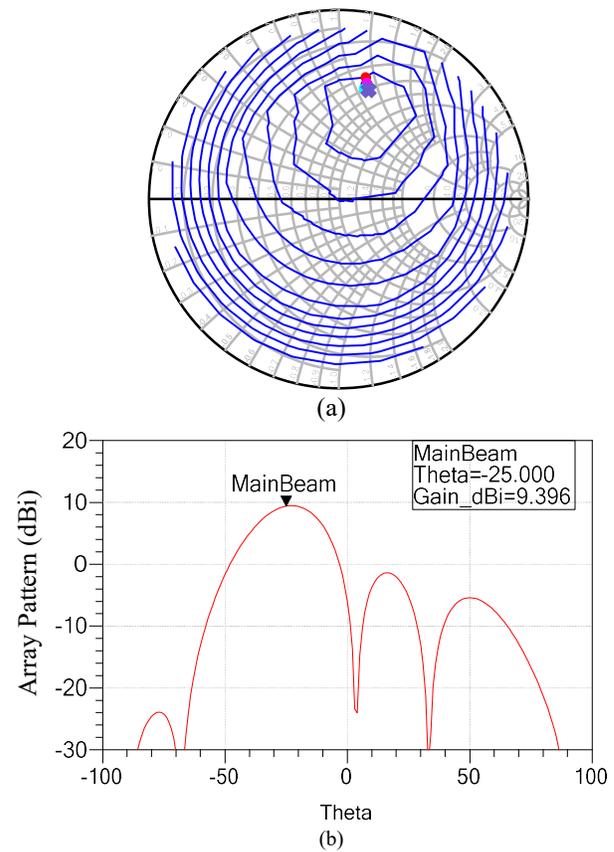

Fig. 16. (a) Tuned driven element impedances at $\theta_s = -25°$, (b) tuned array pattern (dBi).

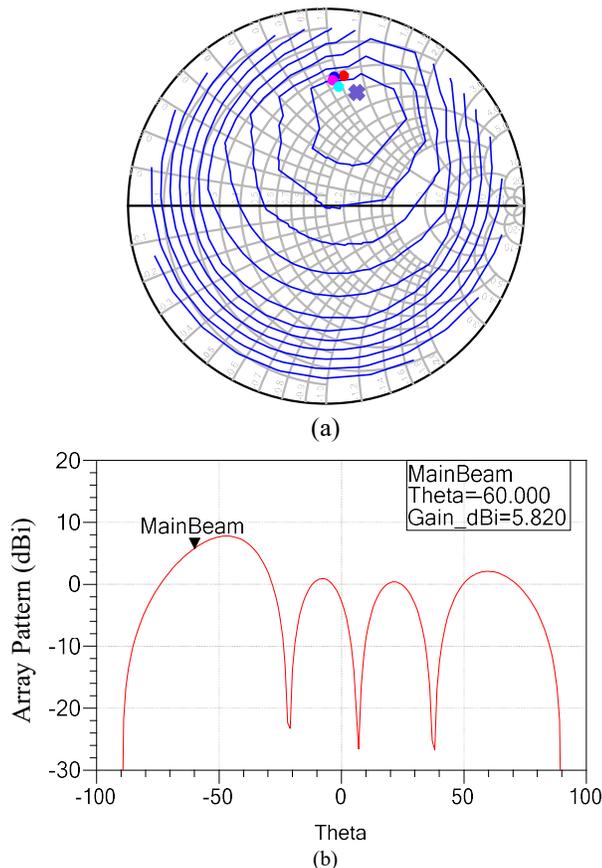

Fig. 15. (a) Tuned driven element impedances at $\theta_s = -60°$, (b) tuned array gain pattern (dBi).

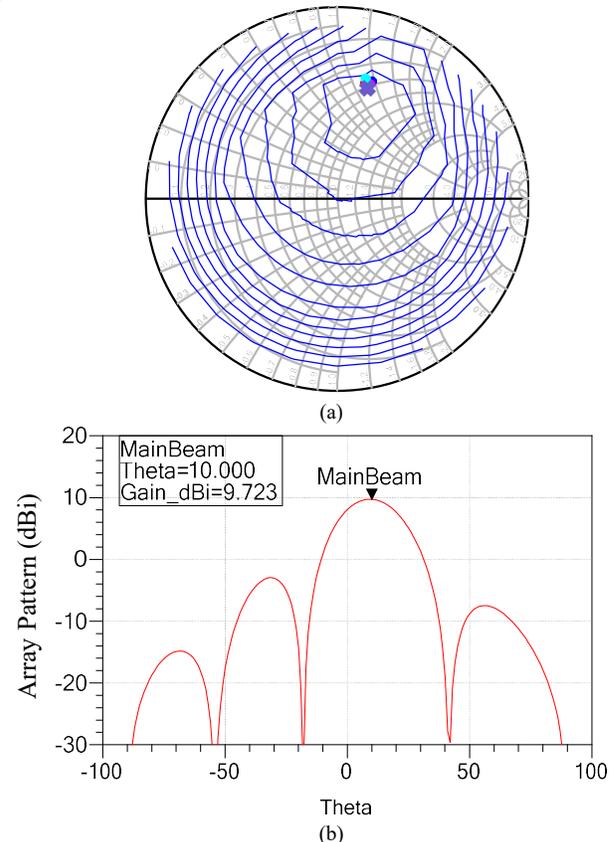

Fig. 17. (a) Tuned driven element impedances at $\theta_s = +10°$, (b) tuned array pattern (dBi).

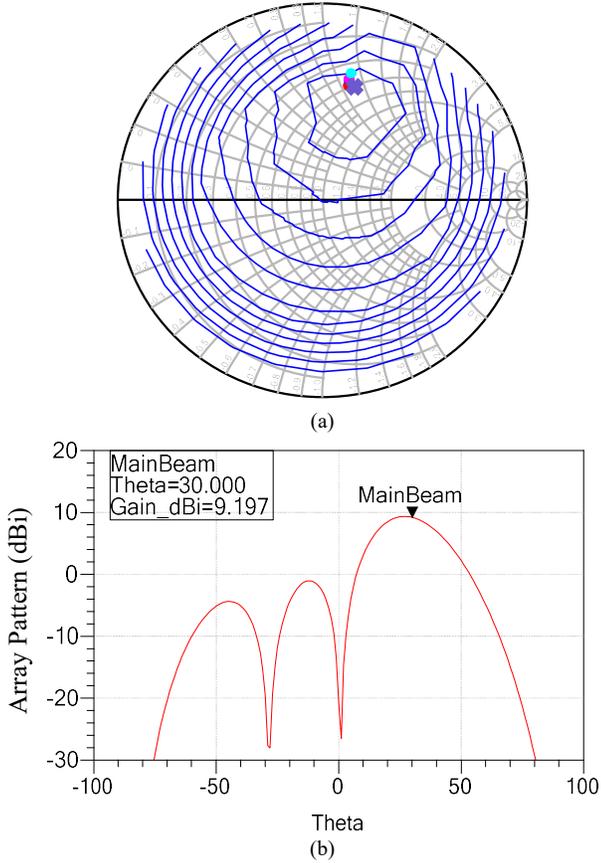

(a)

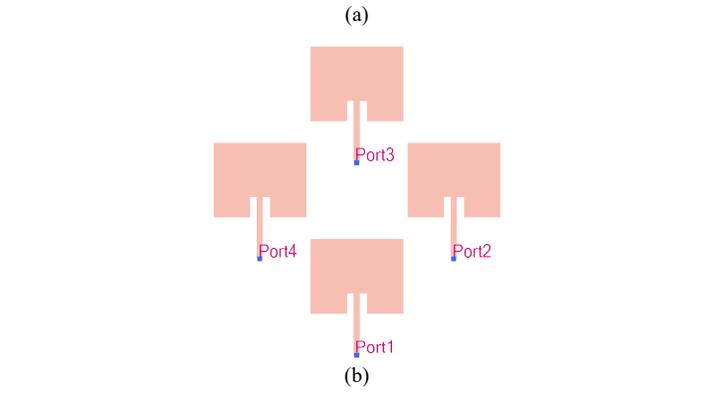

(a)

(b)

Fig. 19. (a) 4-element $\lambda/2$-spaced uniform rectangular microstrip array in ADS, (b) 4-element $\lambda/2$-radius uniform circular microstrip array in ADS.

The first row of elements is excited the same as the second row of elements. The linear phase progressions for the URA are maintained across the first row and the phase progression is reset and maintained across the second row of elements.

The individual current sources for the UCA shown in Fig. 19(b) are assigned the following phase progression [12],

$$I_1 = |I_1| e^{-jk_0 a \sin\theta_s \cos(\phi_s - \phi_1)}$$
$$I_2 = |I_2| e^{-jk_0 a \sin\theta_s \cos(\phi_s - \phi_2)}$$
$$I_3 = |I_3| e^{-jk_0 a \sin\theta_s \cos(\phi_s - \phi_3)}$$
$$I_4 = |I_4| e^{-jk_0 a \sin\theta_s \cos(\phi_s - \phi_4)}. \quad (12)$$

The phase progressions in the UCA depend on the geometric location $\phi_n$ of the individual antenna element. For a four-element UCA, the elements are placed 90° from one another and typically form a $\lambda/2$ radius $a$ around a central circular ring. Fig. 20 shows a parameter sweep of the driven element impedances at the 3.55 GHz design frequency as the array scans from $\theta_s$ = -60° to +60° in the $\phi_s$ = 0° cut for both the URA and the UCA that was conducted to observe the mismatch effects of varying the scan angle and array geometry.

The mutual coupling effects (7) slightly vary in the URA and the UCA compared to the ULA because of the different current excitation phase progression configurations needed to steer the different arrays to the desired scan angle (11)-(12). Nonetheless, the switch-state stubs can once again be used to tune the initial driven impedances to the desired PA load impedance. Figs. 21 and 22 show the initial driven impedances of the URA and UCA, respectively, when steered to broadside.

Fig. 18. (a) Tuned driven element impedances at $\theta_s = +30°$, (b) tuned array pattern (dBi).

## C. Tuning in Varying Array Geometries

Element-wise impedance tuning is not limited to uniformly spaced linear arrays. Fig. 20 shows modified array designs for a uniform rectangular array (URA, Fig. 20(a)) and a uniform circular array (UCA, Fig. 20(b)). The different array configurations have different mutual coupling effects due to the modified current source phase progressions needed to steer the respective arrays.

Equation (7) can be used to calculate the driven impedances. The different current-source phase progressions cause slightly different mutual coupling effects; however, keeping these phase progressions while tuning the driven element impedances is still applicable to maintain the shape and steering angle of the respective array patterns. To steer the beam to a direction $\theta_s$, the individual current sources for the URA shown in Fig. 19(a) are assigned the following phase progression [12]:

$$I_1 = |I_1| \qquad I_2 = |I_2| e^{-jk_0 d \sin\theta_s},$$
$$I_3 = |I_3| \qquad I_4 = |I_4| e^{-jk_0 d \sin\theta_s}. \quad (11)$$

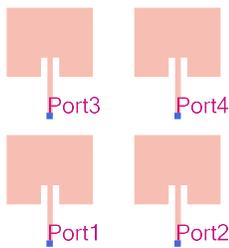

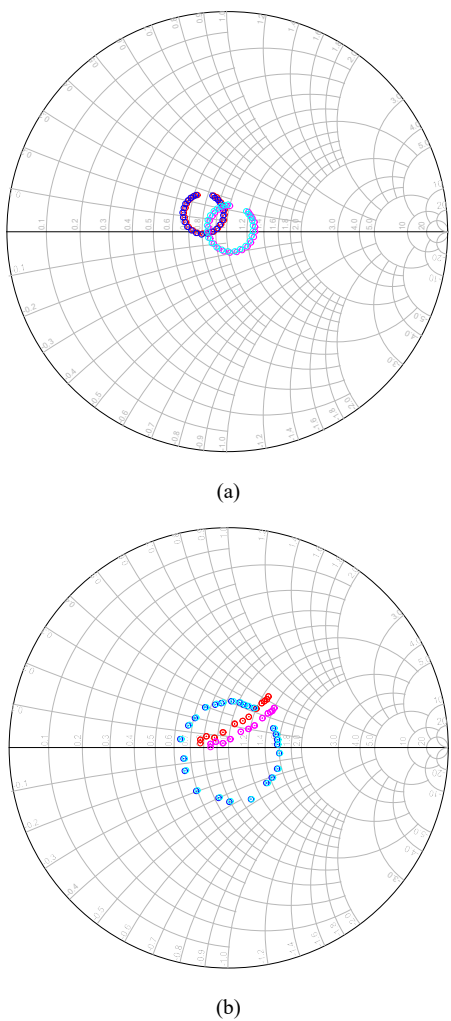

(a)

(b)

Fig. 20. (a) URA parameter swept driven element impedances for scan angles of $\theta_s = -60°$ to $+60°$ at 3.55 GHz for $k_0 = 2\pi/\lambda$, $d = \lambda/2$, (b) UCA parameter swept driven element impedances for scan angles of $\theta_s = -60°$ to $+60°$ at 3.55 GHz for $k_0 = 2\pi/\lambda$, $a = \lambda/2$. End elements are shown in light blue and red, and inner elements are shown in dark blue and purple.

The difference in driven element impedances, due to differences in mutual coupling effects, for the two geometric array configurations is evident in Fig. 21(a) and Fig. 22(a). The switch state stub tuner, however, can overcome these effects, no matter the geometric array configuration, and tune all driven element impedances to the desired maximum power delivered PA load impedance, as shown in Figs. 23 and 24 for the URA and UCA, respectively. Implementing the iterative algorithm once again provides the tuner switch state of 111110 that matches the driven element impedances to the optimum PA load impedance with the lowest loss; maximizing the power fed to the array elements. Table IV shows the range improvement results of the switched state stub tuner when steering the URA and UCA to broadside. Each tuner was set to the same switch state of 111110 and the results show that the array pattern shape and scan angle can be maintained when tuning each individual element with the same switch state while simultaneously achieving a significant improvement in radar range capabilities.

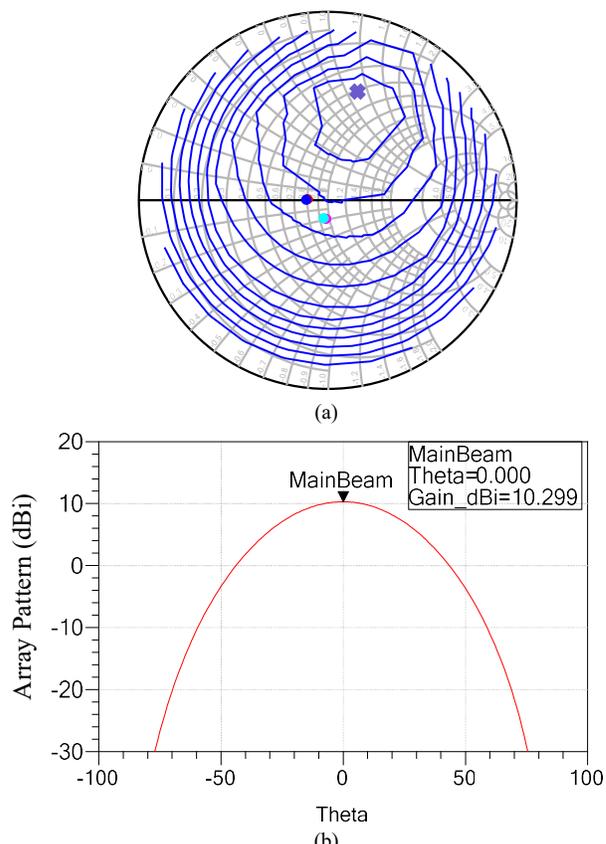

(a)

(b)

Fig. 21. (a) URA untuned driven element impedances for $\theta_s = 0°$, (b) untuned URA array pattern (dBi)

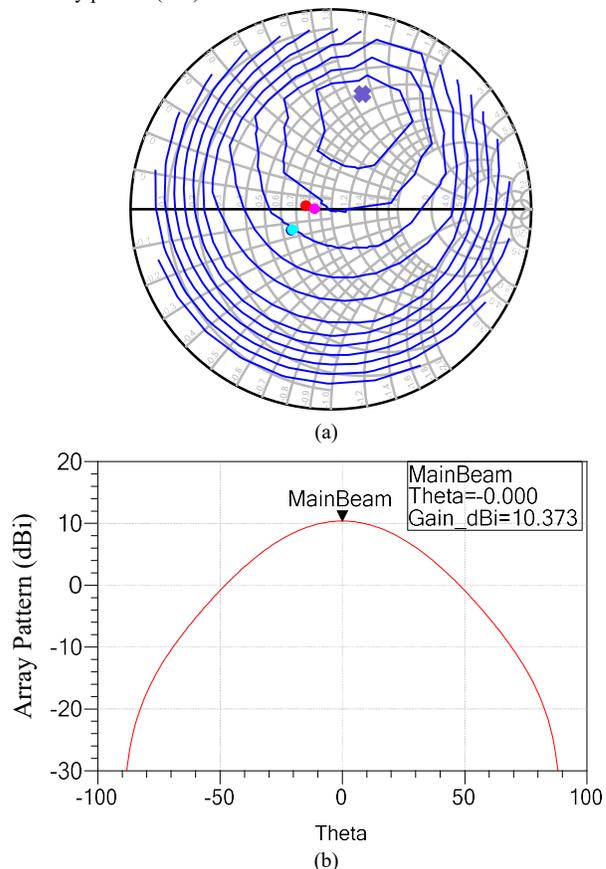

(a)

(b)

Fig. 22. (a) UCA untuned driven element impedances for $\theta_s = 0°$, (b) untuned UCA array pattern (dBi)

## IV. CONCLUSION

The effects of impedance tuning in the individual elements of a phased array transmitter on detection range capabilities have been demonstrated in a joint circuit and electromagnetic simulation platform. For an EM simulation of four-element, $\lambda/2$ spaced, microstrip linear, rectangular, and circular arrays, an increase in output power and calculated radar range is achieved through element-wise impedance tuning. Comparison of the tuned array to an un-tuned array results in calculated radar range increases of over 12 percent in our simulation results, and often over 20 percent for some scan angles. While the output power is increased significantly, very little effect on the relative transmission array pattern is observed if the individual elements are tuned identically, because identical tuning preserves the relative magnitude ratios and phase shifts of the element antenna currents. This experiment demonstrates the potential benefits obtainable by placing real-time impedance tuning in the elements of phased array transmitters for radar, with the results extending to fifth-generation (5G) directional wireless communication applications utilizing phased arrays.

## ACKNOWLEDGMENTS

The authors wish to thank Keysight Technologies for donation of the Advanced Design System software to Baylor University.

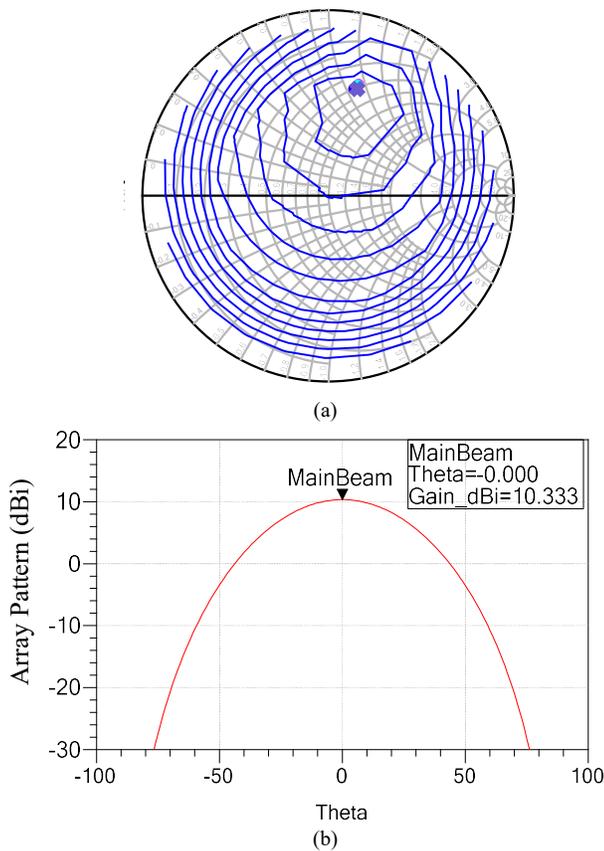

Fig. 23. (a) URA tuned driven element impedances for $\theta_s = 0°$, (b) identically tuned URA array gain pattern (dBi).

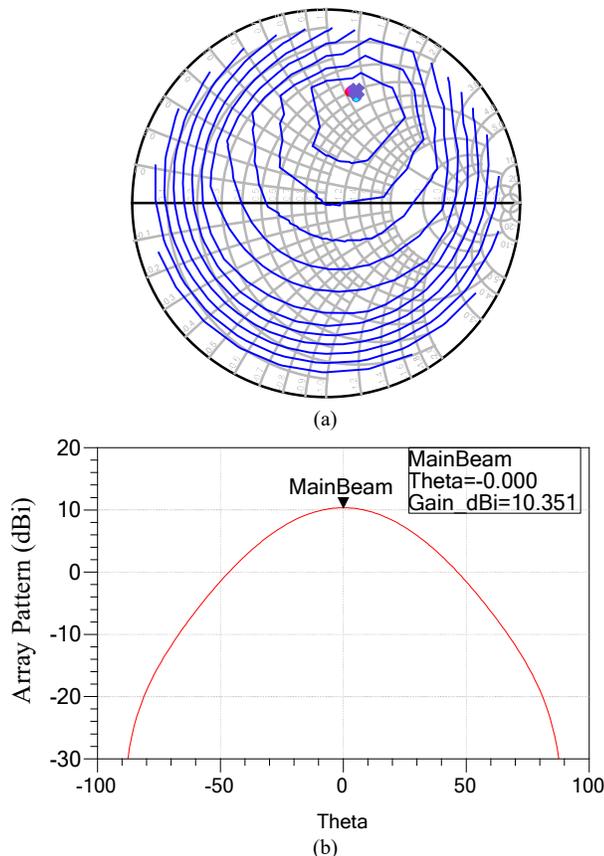

Fig. 24. (a) UCA tuned driven element impedances for $\theta_s = 0°$, (b) identically tuned UCA array gain pattern (dBi).